\documentclass{raa}           
\usepackage{graphicx,times}
\usepackage{natbib}
\usepackage{multirow}
\usepackage{amssymb,amsmath}

\bibpunct{(}{)}{;}{a}{}{,}
\usepackage[a4paper=true,pagebackref=true]{hyperref}
\hypersetup{colorlinks = true, linkcolor = blue, anchorcolor = red, citecolor = blue, filecolor = red, pagecolor = red, urlcolor = red}

\begin{document}
	\title{On the Magnetic Fields of Ultraluminous X-ray Pulsars}
	\volnopage{ {\bf 20XX} Vol.\ {\bf X} No. {\bf XX}, 000--000}
	\setcounter{page}{1}
	\author{Shi-Jie Gao\inst{1,2}, Xiang-Dong Li\inst{1,2}}
	\institute{
		\inst{1} School of Astronomy and Space Science, Nanjing University, Nanjing 210023, China; \textit{lixd@nju.edu.cn}\\
		\inst{2} Key Laboratory of Modern Astronomy and Astrophysics, Nanjing University, Ministry of Education, Nanjing 210023, China\\
		\vs \no
		{\small Received 20XX Month Day; accepted 20XX Month Day}}
	\abstract{		
		 So far quite a few ultraluminous X-ray (ULX) pulsars have been discovered. In this work, we construct a super-Eddington, magnetic accretion disk model to estimate the dipole magnetic field of eight ULX pulsars based on their observed spin-up variations and luminosities. We obtain two branches of dipole magnetic field solutions. They are distributed in the range of $B\sim (0.16-64.5)\times 10^{10}\,{\rm G}$ and $\sim (0.275-79.0)\times 10^{13}\,{\rm G}$ corresponding to the low- and high-$B$ solutions respectively. The low magnetic field solutions correspond to the state that the neutron stars are far away from the spin equilibrium, and the high magnetic field solutions are close to the spin equilibrium. The ultra-strong magnetic fields derived in Be-type ULX pulsars imply that the accretion mode in Be-type ULX pulsars could be more complicated than in the persistent ULX pulsars and may not be accounted for by the magnetized accretion disk model. We suggest that the transition between the accretor and the propeller regimes may be used to distinguish between the low- and high-$B$ magnetic field solutions in addition to the detection of the cyclotron resonance scattering features.
		 }
	 
	\authorrunning{S.-J. Gao \& X.-D. Li }
	\titlerunning{On the Magnetic Fields of ULX Pulsars}
	\maketitle
	
	\section{Introduction}
	\hspace{15pt}Ultraluminous X-ray sources (ULXs) are extranuclear X-ray sources with isotropic luminosity exceeding $\sim{10}^{39}\,{\rm erg\ s^{-1}}$, the Eddington limit ($L_{\rm Edd}$) for a standard stellar-mass accretor (\citealt{ULXs+2017+ARA&A}). ULXs are usually considered to be rapidly accreting steller-mass black holes and neutron stars (NS) in X-ray binaries, while \cite{Colbert+1999+intermediatemassBH} suggested that some ULXs may contain intermediate-mass black holes accreting at a sub-Eddington rate. Recently, the discovery of pulsations in M82 X-2 (\citealt{Bachetti+m82x2+2014+nature}) unveils that M82 X-2 is powered by an accreting NS rather than a black hole. \cite{2015+Shao+ULXspopulation} showed that NS ULXs may significantly contribute to the ULX population compared with black hole X-ray binaries using binary population synthesis and detailed binary evolution calculations. \cite{King+2017+UlXs} suggested that many unpulsed ULXs may actually contain NSs rather than black holes.
	
	According to their accretion features, there are two types of ULX pulsars, the persistent ones with (possible) OB supergiant companions such as M82 X-2 (\citealt{Bachetti+m82x2+2014+nature}), NGC5907 ULX-1 (\citealt{Israel+2017+NGC5907ulx1}), NGC5907 ULX-1  (\citealt{Israel+2017+NGC5907ulx1}), M51 ULX-7 (\citealt{M51ulx7+Rodrguez+2020ApJ}) and  NGC7793 P13  (\citealt{NGC7793p13+Furst+2016,Furst+2018,Israel+2017+NGC7793P13}); and the transient ones with Be star companions such as NGC300 ULX-1 (\citealt{Carpano+2018}), SMC X-3 (\citealt{Tsygankov+2016MNRAS.457.1101T,Townsend+2017}), NGC2403 ULX (\citealt{Trudolyubov+2007ApJ}) and  Swift J0234.6+6124 (\citealt{Doroshenko+swift+2018,Swift+2018+distance}). In both cases, the NSs are thought to be accreting via an accretion disk.
	
	The dipole magnetic field strength plays an important role in the nature and evolution of NS ULXs. There are several ways to estimate their dipole magnetic field strengths. One is using the observed spin variation to derive the magnitude of the accretion torque which depends on the interaction between the NS dipole magnetic field and the accretion disk (\citealt{GL79a,GL79b}). Another method is identifying the critical state in which the magnetospheric radius equals the co-rotation radius when a transition between the accretion and propeller regimes occurs (eg., \citealt{Tsygankov+2016MNRAS.457.1101T,Tsygankov+2017}). The third and more accurate one is calculating the magnetic field from the cyclotron resonance scattering features (CRSFs) detected in the energy spectrum caused by transition of charged particles between different quantum Landau levels (\citealt{Walter+2015+A&ARv+HMXB}). \cite{Brightman+2018+nature+cyc} detected an absorption line at $4.5\,\rm{keV}$ in the \textit{Chandra} spectrum of M51 ULX-8, which implies a magnetic field of $\sim 10^{11} \,\rm{G}$ or $\sim 10^{15} \,\rm{G}$ corresponding to scattering off electrons or protons respectively. However, \cite{Middleton+MNRAS+2019+cyc} reanalyzed the data and gave an upper limit on the dipole magnetic field of $10^{12}\,\rm{G}$ and ruled out a very strong ($10^{15}\,{\rm G}$) dipole magnetic field.
	
	There have been many theoretical works on the nature of the ULX pulsars. Using a torque model of \cite{GL79b}, \cite{DallOsso+etl+2015} studied the magnetic field of M82 X-2 and obtained three kinds of magnetic field solutions, corresponding to $5\times10^{9}-10^{11}\,{\rm G}$, $10^{11}-10^{12}\,{\rm G}$, and $10^{12}-10^{13}\,{\rm G}$, respectively. Although it is hard to distinguish which one is more reasonable, \cite{DallOsso+etl+2015} prefer the highest one ($\sim 10^{13}\,{\rm G}$) which implies that M82 X-2 is close to the spin equilibrium state, able to account for the fluctuations of the spin-up rate and the spin reversion. Meanwhile, a high magnetic field can reduce the electron scattering cross-section and enhance the maximum accretion luminosity. \cite{Erkut+2020ApJ...899...97E} systematically studied the magnetic fields of known ULX pulsars and showed that the magnetic fields are in the range of $10^{11}-10^{15}\,{\rm G}$. But they argued that it is not necessary for the pulsars to have magnetar-strength fields if radiative beaming is taken into consideration. Other works for example, \cite{M82X-2+Eksi+2015},  \cite{Xu+Li+2017+M82x-2}, \cite{King+2017+UlXs}, \cite{King2019+no+magnetar}, and \cite{M51ULX7+Vasilopoulos+2020} also studied the magnetic fields of the ULX pulsars. In these works, the torque acting on the NSs was usually derived based on the \cite{GL79a,GL79b} magnetized, Keplerian disk model. However, to construct the accretion torque model for ULX pulsars, one needs to adequately consider the effects of super-Eddington accretion and disk dynamics.
	
	In this paper, we derive the magnetic fields of eight ULX pulsars from their spin evolution. In Section~\ref{section:models}, we construct an accretion torque model taking into account the NS-accretion disk interaction and mass loss for a super-Eddington accretion disk. Eight ULX pulsars, M82 X-2, NGC5907 ULX-1, M51 ULX-7, NGC7793 P13, NGC300 ULX-1, SMC X-3, NGC2403 ULX and Swift J0234.6+6124 are studied and we calculate their magnetic fields in Section~\ref{section:results}. Finally, we discuss possible observational implications of the results and summarize our work in Section~\ref{section:disscussion}.
	
	\section{model}
	\label{section:models}
		
	\hspace{15pt}Our model is based on the work of \cite{GL79a,GL79b} and \cite{Wang+1987A&A...183..257W,Wang+1995}. In this model, the NS magnetic field disrupts the disk flow inside the inner radius of the disk $R_{0}$. There exists a co-rotation radius $R_{\rm c}=(GM/\Omega_{\rm s}^2)^{1/3}$ at which the Keplerian angular velocity $\Omega_{\rm K}$ of the plasma in the disk equals the spin angular velocity $\Omega_{\rm s}$ of the NS, where $G$ is the gravitational constant and $M$ is the mass of NS. If $R_0<R_{\rm c}$, stable accretion occurs and the matter is transferred to the NS following the magnetic field lines around $R_0$. If $R_0>R_{\rm c}$, the NS enters the propeller regime where the accreted matter is ejected from the NS because the centrifugal force is greater than the gravitational force. We assume that the NSs have a magnetic field with its axis aligned with the spin axis and perpendicular to the accretion disk. Here we introduce a cylindrical coordinate system $\left(R,\phi,z\right)$ centred on the NS, and in the case of steady accretion, the transfer of angular momentum generates a torque
	\begin{equation}
		\label{eq:N_0}
		N_0{\simeq{\dot{M}}_{\rm in} R_0^2\Omega_{\rm K}(R_0)}=\dot M_{\rm in}(GMR_0)^{1/2},
	\end{equation}
	where $\dot{M}_{\rm in}$ denotes the accretion rate at the inner radius of the accretion disk. As we will mention below, the rotational behavior of the disk matter at the inner edge of the disk deviates from Keplerian rotation, but we assume that the magnitude of the angular velocity at $R_0$ is very close to its Keplerian value. On the accretion disk surface, the dipole magnetic field component of the NS in the $z$-direction is
	\begin{equation}
		\label{eq:B_z}
		B_z=-\eta\frac{\mu}{R^3},
	\end{equation}
	where $\eta \le 1$ is a screening coefficient and is usually taken as unity (\citealt{GL79a,GL79b,Livio+1992MNRAS.259P..23L}), and $\mu=BR_{\rm NS}^3$ denotes the NS’s magnetic moment, where $R_{\rm NS}$ is the radius of the NS.
	
	We adopt the magnetically threaded disk (MTD) model (\citealt{GL79a,GL79b,Wang+1995}) to describe the mechanism of the interaction between the magnetic field of NS and the accretion disk. In the MTD model, the magnetic field lines of an NS penetrate the accretion disk and are distorted due to the shearing motion between the differential rotation of the accretion disk and the spin of the NS, generating a toroidal component of the magnetic field in the $\phi$-direction
	\begin{equation}
		\label{eq:B_phi}
		\frac{B_{\rm\phi}}{\tau_{\rm \phi}}=\gamma\left(\Omega_{\rm s}-\Omega_{\rm K}\right)B_{z},
	\end{equation}
	where $\tau_{\rm \phi}$ denotes the dissipation timescale for $B_{\rm \phi }$ and $\gamma \gtrsim 1$ is a numerical factor which depends on the steepness of the transition between the Keplerian motion inside the disk and co-rotation with the star outside the disk (\citealt{GL79a,GL79b,Wang+1995}). The torque generated by the interaction between the magnetic field and the disk is
	\begin{equation}
		\label{eq:N_mag}
		N_{\rm mag}=-\int_{R_0}^{\infty}{B_{z}B_{\rm \phi } R^2\mathrm d R}.
	\end{equation}
	
	The total torque on the NS can be expressed in the dimensionless form
	\begin{equation}
		\label{eq:n_total}
		n(\omega)=\frac{N_0+N_{\rm mag}}{N_0},
	\end{equation}
	where $\omega=\Omega_{\rm s}/\Omega_{\rm K}(R_0)=(R_0/R_{\rm c})^{3/2}$ is the fastness parameter. \cite{Wang+1995} derived the function $n(\omega)$, taking into account different forms of $\tau_\phi$ determined by the Alfv\'en speed, turbulent diffusion in the disk and magnetic reconnection outside the disk. He adopted the following boundary condition to determine the inner radius $R_0$ at which the angular momentum that the magnetic field removes from the disk and the internal viscous stress reach balance, that is
	\begin{equation}
		\label{eq:boundary}
		-B_{\rm \phi 0} B_{z0} R_0^2=\left.{\dot{M}}_{\rm in}\frac{\mathrm d}{\mathrm d R}\left(\Omega R^2\right)\right|_{R=R_0}={\dot{M}}_{\rm in}\left[R^2\frac{\mathrm d\Omega}{\mathrm d R}+2R\Omega\right]_{R=R_0},
	\end{equation}
	where the subscript 0 denotes quantities evaluated at $R=R_0$, and $\Omega$ is the angular velocity of the disk. \cite{Wang+1995} regarded the angular velocity of the disk {around} $R_0$ as Keplerian, i.e., $\Omega(R)=\Omega_{\rm K}(R)=(GM/R^3)^{1/2}$. Substitute it into Equation~(\ref{eq:boundary}), and it follows
	\begin{equation}
		\label{eq:boundary_wang}
		\frac{B_{\rm \phi 0} B_{z0}}{\dot M_{\rm in}(GMR_0)^{1/2}}=-\frac12\frac{1}{R_0^3}.
	\end{equation}

	Because the dynamical viscosity is sufficiently small and the magnetic stress is dominant and greater than the shear stress around the inner edge of the disk, the rotational behaviour of the disk matter around $R_0$ deviates from Keplerian and the disk matter is forced to corotate with the spin of the NS inside $R_0$. In our study, we adopt the rotational behaviour of the accretion disk at the inner edge suggested by \cite{Li+Wang+1996+1996A&A...307L...5L} rather than Keplerian, that is, at the inner radius of the disk, the angular velocity $\Omega(R_{0})$ of the disk reaches its maximum value and begins to deviate from $\Omega_{\rm K}(R_0)$, so, $\Omega(R_0)\simeq\Omega_{\rm K}(R_0)$ and $\left.\frac{\mathrm d\Omega}{\mathrm d R}\right|_{R=R_0}=0$ ({see fig.~1 in} \citealt{Li+Wang+1996+1996A&A...307L...5L}). Some analytical calculations and magnetohydrodynamics simulations support this hypothesis, for example, \cite{Erkut+2004ApJ...617..461E}, \cite{Long2005}, \cite{Romanova+2008ApJ...673L.171R},  \cite{Zanni+2009A&A...508.1117Z,Zanni+2013A&A...550A..99Z} and \cite{Faghei+2018MNRAS.473.2822F}. From Equation~(\ref{eq:boundary}) we have
	\begin{equation}
		\label{eq:boundary_our}
		\frac{B_{\rm \phi 0} B_{z0}}{\dot M_{\rm in}(GMR_0)^{1/2}}=-2\frac{1}{R_0^3}.
	\end{equation}
	 It should be noted that the left hand side of the Equation~(\ref{eq:boundary_our}) is four times that of the Equation~(\ref{eq:boundary_wang}). 
	
	We use the relation between the magnetic field components (\citealt{Wang+1995}, eq.~13; hereafter, case~1), i.e.,
	\begin{equation}
		\label{eq:B_z_B_phi_case1}
		\frac{B_\phi}{B_{z}}=\frac{\gamma}{\alpha}\frac{\Omega_{\rm s}-\Omega_{\rm K}}{\Omega_{\rm K}},
	\end{equation}
	where $\alpha$ is a numerical factor less than unity (\citealt{Wang+1995}). In this case, turbulent mixing within the disk limits the growth of $B_{\phi}$ and the velocity $v_{\rm t}$ of the dominant turbulent eddies scales as the sound speed $c_{\rm s}$, so the dissipation timescale of $B_{\phi}$ is $\tau_{\phi}=(\alpha\Omega_{\rm K})^{-1}$. When $R=R_0$, we have $B_{\phi0}/B_{z0}=-\gamma(1-\omega)/\alpha$ and $B_{z0}=\eta\mu R_0^{-3}$. Combining this with Equation~(\ref{eq:boundary_our}) to eliminate $B_{\phi0}$ and $B_{z0}$, the total dimensionless torque of Equation~(\ref{eq:n_total}) can be derived to be
	\begin{equation}
		n(\omega)=1+\frac{2}{3}\,\frac{1-2\omega}{1-\omega}.
	\end{equation}	
	We also consider another form of the magnetic field relation (\citealt{Wang+1995}, eq.~17; hereafter, case~2), that is 
	\begin{equation}
		\frac{B_{\phi}}{B_{z}}=
		\begin{cases}
			{\gamma_{\rm max} (\Omega_{\rm s}-\Omega_{\rm K})/\Omega_{\rm K},}&{R\leq R_{\rm c}}\\
			{\gamma_{\rm max} (\Omega_{\rm s}-\Omega_{\rm K})/\Omega_{\rm s},}&{R\geq R_{\rm c}}
		\end{cases}, 
	\end{equation}
	where $\gamma_{\rm max}\sim 1$ is the maximum value limited by magnetic reconnection taking place outside the disk. In this case, the magnetic field lines continually rearrange their connections to the disk to balance the magnetospheric stresses and the shearing motion occurs on the same timescale. Similar as in case~1, we obtain 
	\begin{equation}
		n(\omega)=1+\frac{2}{9}\,\frac{2\omega^2-6\omega+3}{1-\omega}.
	\end{equation}	
	
	\begin{figure}[h]
		\begin{subfigure}{.5\textwidth}
			\includegraphics[width=\textwidth]{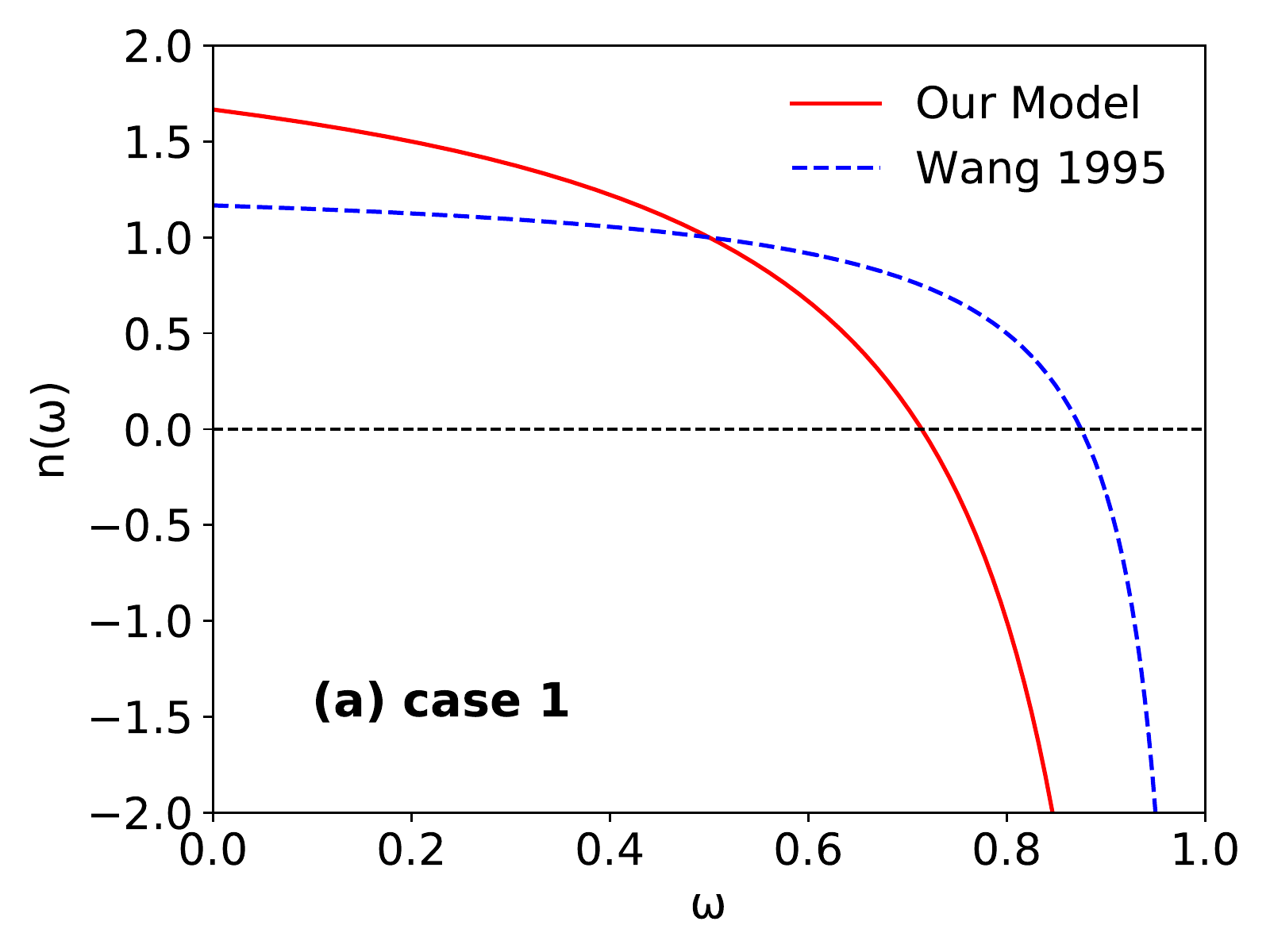}
		\end{subfigure}
		\begin{subfigure}{.5\textwidth}
			\includegraphics[width=\textwidth]{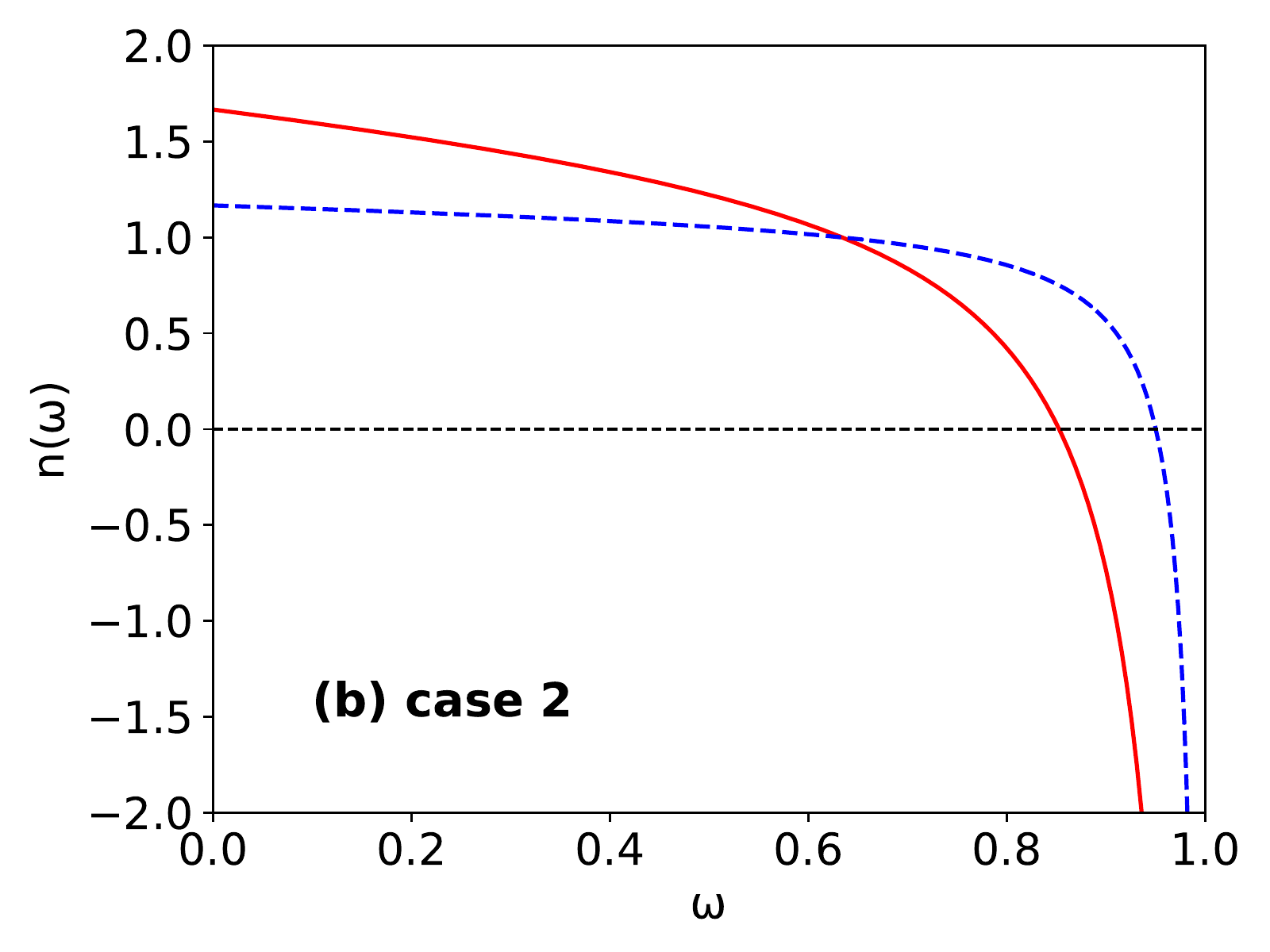}
		\end{subfigure}
		\caption{\small The relation between $n(\omega)$ and $\omega$, panel (a) and (b) represent cases~1 and~2 respectively. The red solid lines and blue dashed lines represent the relation in our model and in \cite{Wang+1995} model, respectively.}
		\label{fig:dimensionless_n}
	\end{figure}
	Figure~\ref{fig:dimensionless_n} shows the relation between the dimensionless torque $n$ and the fastness parameter $\omega$, and panels (a) and (b) represent cases~1 and~2 respectively. From Figure~\ref{fig:dimensionless_n}, we find that $n(\omega)$ decreases with increasing $\omega$. As a result, there is an equilibrium state where $n(\omega)$ equals $0$ and the critical fastness parameter $\omega_{\rm crit} \simeq 0.714$ and $0.853$ compared with larger values $\omega_{\rm crit}\simeq 0.875$ and $0.967$ derived by \cite{Wang+1995} for cases 1 and 2, respectively.
	
	The inner radius $R_0$ of the disk can also be expressed in the following form,
	\begin{equation}
		\label{eq:R_0}
		R_0=\xi R_{\rm A},
	\end{equation}
	where $\xi$ is usually a constant of order of unity which relates the inner radius of the disk with the Alfv\'en radius $R_A=(\mu^4/2GM{\dot{M}^2}_{\rm in})^{1/7}$ at which the magnetic pressure ($B^2/8\pi$) and the ram pressure balance for spherical accretion (\citealt{Accretionpower+2002apa..book.....F}). The implication of Equation~(\ref{eq:R_0}) is that the inner boundary condition should satisfy both torque balance and pressure balance. \cite{GL79a,GL79b} obtained $\xi=0.52$ from numerical calculations; \cite{Wang1996} pointed out that $\xi$ is determined by the fraction of the star's magnetic flux threading the disk and is usually $\xi\simeq 1$; \cite{Long2005} used axisymmetric magnetohydrodynamics simulations to investigate the equilibrium state of accretion rotating magnetic stars and obtained $\xi\simeq 0.5$, considering a relatively weak magnetic field with a high coronal density and a stronger magnetic field with a lower coronal density respectively; \cite{Kulkarni20073D} performed 3D simulations of magnetospheric accretion and gave $\xi\sim 0.55-0.72$. From Equation~(\ref{eq:boundary_our}) and (\ref{eq:R_0}), we obtain for case~1,
	\begin{equation}
		\xi=2^{-1/7}\eta^{4/7}\gamma^{2/7}\alpha^{-2/7}(1-\omega)^{2/7},
	\end{equation}
	and similarly, in case~2,
	\begin{equation}
	\xi=2^{-1/7}\eta^{4/7}\gamma_{\rm max}^{2/7}(1-\omega)^{2/7}.
	\end{equation}
	{Because the reconnection of the field lines taking place at the inner radius of the disk prevents the magnetic pitch $B_{\phi 0}/B_{\rm z0}$ exceeding unity (\citealt{Wang1996}), we take $\eta=1$, $\gamma/\alpha\simeq1$ and $\gamma_{\rm max}\simeq1$.} Therefore, $\xi\simeq2^{-1/7}(1-\omega)^{2/7}$ for both cases 1 and 2.
	
	The above derivation of the torque is limited to the model of sub-Eddington accretion disks. \cite{SS1973} investigated the structure of accretion disks when the accretion luminosity is higher than the local Eddington luminosity $L_{\rm Edd}$. They pointed out that the accretion rate in the disk can keep invariant outside the spherization radius $R_{\rm sph}={3GM\dot{M}}/{2L_{\rm Edd}}$ at which the Eddington-limited accretion occurs, and is Eddington-limited within $R_{\rm sph}$. Thus, depending on whether the inner disk radius is inside and outside the spherization radius, the accretion rate at the inner disk radius can be expressed to be (e.g., \citealt{Xu+2019+mdot_mdot_in,Erkut+2020ApJ...899...97E}),
	\begin{equation}
		\dot{M}_{\rm in}=
		\begin{cases}
			{\dot{M},}& {R_0 \geq R_{\rm sph}}\\
			{\dot{M} R_0/R_{\rm sph},}&{R_0 \leq R_{\rm sph}}
		\end{cases}.
	\label{eq:Mdotin}
	\end{equation}
	We can see that when $R_0\leq R_{\rm sph}$, the inner radius of the disk can be expressed as 
	\begin{equation}
		\label{eq:R0_no_mdot}
		R_0=\xi^{7/9}\left(\frac{\mu^4}{2GM}\right)^{1/9}\left(\frac{2L_{\rm Edd}}{3GM}\right)^{-2/9},
	\end{equation}
	and the mass transfer rate  at the inner radius of the disk is
	\begin{equation}
		{\dot{M}}_{\rm in}=2^{2/3}3^{-7/9}\xi^{7/9}(GM)^{-8/9}L_{\rm Edd}^{7/9}{\mu}^{4/9}.
	\end{equation}
	It is found that, taking $\xi$ as a constant, $R_0$ and ${\dot{M}}_{\rm in}$ no longer contain $\dot{M}$, if $R_0 \leq R_{\rm sph}$. For super-Eddington accretion, radiation is not isotropic but collimated, which complicates the relation between the mass transfer rate and the isotropic (apparent) luminosity, Following \cite{King+2009+ULXs}, \cite{King16+NSvsBH} and \cite{King+2017+UlXs}, we use the following relation between the isotropic luminosity and the accretion rate
	\begin{equation}
		L=
		\begin{cases}
			{\epsilon\dot{m}\dot{M}_{\rm Edd} c^2,}				&{\dot{m}\leq 1}\\
			{L_{\rm Edd}(1+\ln{\dot{m}}),}			&{1\leq \dot{m} \leq \sqrt{73}}\\
			{L_{\rm Edd}(1+\ln{\dot{m}})/b,}			&{\dot{m} \geq \sqrt{73}}\\
		\end{cases},
		\label{eq:L_mdot}
	\end{equation}
	where $\epsilon\sim 0.15$ is the radiation efficiency for NSs, $\dot{m}=\dot{M}/\dot{M}_{\rm Edd}$, $c$ is the velocity of light in vacuum, $b\simeq73/{\dot{m}}^2$ is the beaming factor. To guarantee that the change in the luminosity with $\dot m$ is continuous at $\dot{m}=1$, we take $\dot{M}_{\rm Edd}=1.60\times{10}^{18}\,{\rm g\ s^{-1}}$ and  $L_{\rm Edd}=2.16\times  10^{38}\,{\rm erg\ s^{-1}}$. 
		
	We assume that the NS is rigidly rotating, and its spin evolution is determined by
	\begin{equation}
		-2\pi I \dot{P}/P^2=nN_0,
		\label{eq:roation}
	\end{equation}
	where $I=2M R_{\rm NS}^2/5$ is the moment of inertia. Substituting Equation (\ref{eq:N_0}) into Equation (\ref{eq:roation}) leads to
	\begin{equation}
		-2\pi I \dot{P}/P^2={\dot{M}}_{\rm in}\left(GMR_{\rm c}\right)^{1/2}n(\omega)\omega^{1/3}.
		\label{eq:P,Pdot,omega}
	\end{equation}
	It can be further obtained that if $R_{\rm 0} \geq R_{\rm sph}$,
	\begin{equation}
		\label{eq:solve_omega_geq}
		n(\omega)\omega^{1/3}=-2.478\,M_{1.4}^{1/3}R_{6}^{2}\dot{m}^{-1}\dot{P}_{-10}P_1^{-7/3};
	\end{equation}
	and	if $R_{\rm 0} \leq R_{\rm sph}$,
	\begin{equation}
		\label{eq:solve_omega_leq}
		n(\omega)\omega=-3.057\times10^{-2}\,M_{1.4}R_{\rm 6}^{2}\dot{P}_{-10}P_{1}^{-3},
	\end{equation}
	where $M_{1.4}=M/1.4\rm M_{\odot}$, $R_{6}=R_{\rm NS}/10^{6}\,{\rm cm}$, $\dot{P}_{-10}=\dot{P}/10^{-10}\,{\rm s\ s^{-1}}$ and $P_1=P/1\,{\rm s}$. Using the observed luminosity $L$, spin period $P$ and spin-change rate $\dot{P}$, the fastness parameter $\omega$ can be calculated from Equation~(\ref{eq:solve_omega_geq}) or (\ref{eq:solve_omega_leq}). The left hand side of the Equation~(\ref{eq:solve_omega_geq}) firstly increases and then decreases with increasing $\omega$, and it reaches a maximum value $0.925$ and $0.988$ for cases~1 and~2, respectively. Similarly, the left hand side of the Equation~(\ref{eq:solve_omega_leq}) also firstly increases and then decreases with increasing $\omega$, and it reaches a maximum value $0.506$ and $0.640$ for cases~1 and~2, respectively. As a result, there are solutions of Equation~(\ref{eq:solve_omega_geq}) only when $M_{1.4}^{1/3}R_{6}^{2}\dot{m}^{-1}(-\dot{P}_{-10})P_1^{-7/3}\leq0.37$ and $\leq 0.40$, and of Equation~(\ref{eq:solve_omega_leq}) only when $M_{1.4}R_{\rm 6}^{2}(-\dot{P}_{-10})P_1^{-3}\leq16.5$ and $\leq 20.9$ for cases~1 and~2, respectively. Taking $M_{1.4}=1$ and $R_6=1$, if $0.37<\dot{m}^{-1}(-\dot{P}_{-10})P_1^{-7/3}<0.40$, Equation~(\ref{eq:solve_omega_geq}) is only solvable for case~2, and if $16.5<(-\dot{P}_{-10})P_1^{-3}<20.9$, Equation~(\ref{eq:solve_omega_leq}) is only solvable for case~2. So, there may be at most four solutions of $\omega$ for a given ULX pulsar, two from Equation~(\ref{eq:solve_omega_geq}) and the other two from Equation~(\ref{eq:solve_omega_leq}). If the spin-up rate of a ULX pulsar is too high, there will be no solution in our model, which means that other factors should be considered, for example, wind mass loss from  the magnetosphere caused by the open fields (\citealt{Lovelace+1995MNRAS.275..244L,Romanova+2003ApJ...588..400R}). Using Equation~(\ref{eq:R_0}) and $R_0=\omega^{2/3}R_{\rm c}$, the dipolar magnetic field $B$ of the NS can be obtained, namely, if $R_0\ge R_{\rm sph}$,
	\begin{equation}
		B=1.371\times10^{12}\,\xi^{-7/4}\omega^{7/6}M_{1.4}^{5/6}R_6^{-3}\dot m^{1/2}P_1^{7/6}\,{\rm G};
		\label{eq:B_geq}
	\end{equation}
	and if $R_0\le R_{\rm sph}$,
	\begin{equation}
		B=1.234\times10^{13}\,\xi^{-7/4}\omega^{3/2}M_{1.4}^{1/2}R_6^{-3}P_1^{3/2}\,{\rm G}.
		\label{eq:B_leq}
	\end{equation}
	
	\section{Results}
	\label{section:results}
	
	\hspace{15pt}Our targets are eight ULX pulsars, M82 X-2, NGC5907 ULX-1, M51 ULX-7, NGC7793 P13, NGC300 ULX-1, SMC X-3, NGC2403 ULX and Swift J0234.6+6124. Table~\ref{table:parameters} presents their spin period $P$, spin period derivation $\dot{P}$ and isotropic X-ray luminosity $L$. From their observed X-ray luminosities, we calculate the mass accretion rates using Equations (\ref{eq:L_mdot}) and the beaming factor $b$, which are listed in the fifth and sixth columns of Table~\ref{table:parameters}. We calculate the magnetic field $B$ using Equation~(\ref{eq:solve_omega_geq}) (\ref{eq:solve_omega_leq}) (\ref{eq:B_geq}) and (\ref{eq:B_leq}) based on the observed period $P$, the mass accretion rate $\dot{M}$ and the spin-up rate $\dot P$. In our calculation, we take $M=1.4\,{\rm M_{\odot}}$ and $R_{\rm NS}=10^{6}\,{\rm cm}$.
	
	\begin{figure}[h]
		\begin{subfigure}{.5\textwidth}
			\includegraphics[width=\linewidth]{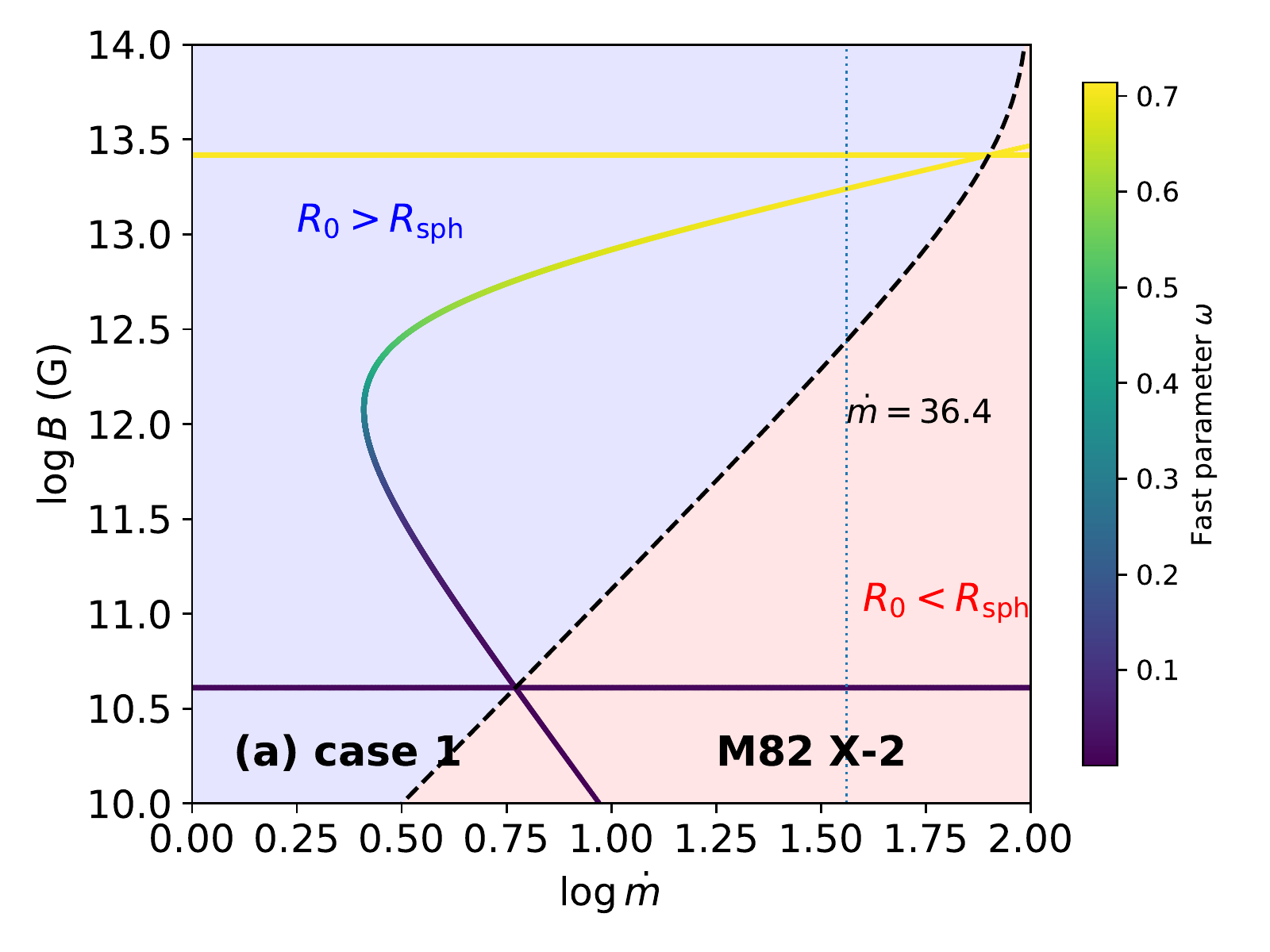}
		\end{subfigure}
		\begin{subfigure}{.5\textwidth}
			\includegraphics[width=\linewidth]{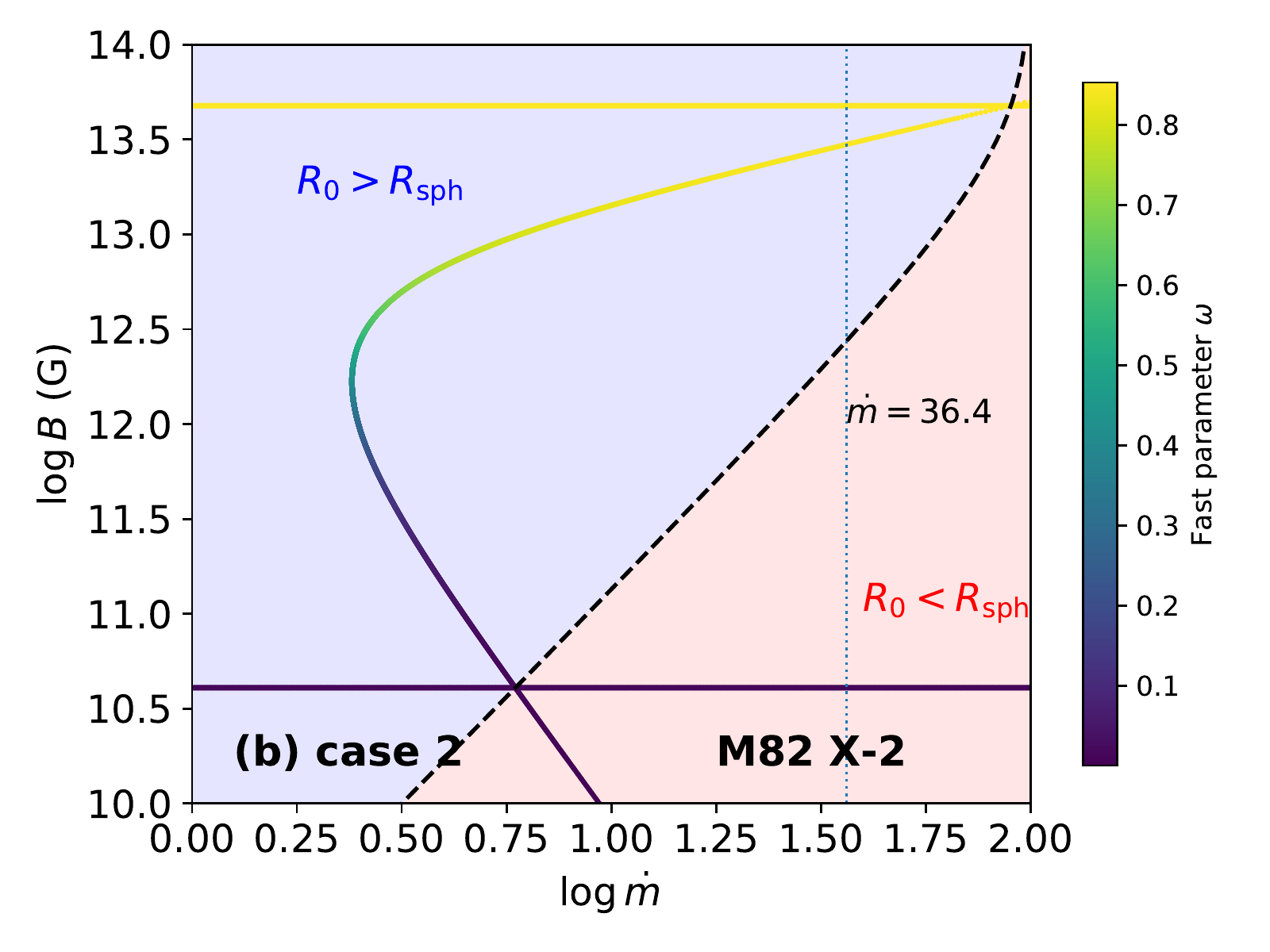}
		\end{subfigure}
		\caption{\small{ The $B-\dot m$ relation with $M_{1.4}=1$, $R_6=1$, $\dot{P}_{-10}=-2$ and $P=1.37\,{\rm s}$ for M82 X-2. The solid curves represent the relation between $\dot m$ and $B$ with different values of $\omega$ which are shown by a color bar. The black dashed line represents $R_0=R_{\rm sph}$, and it divides the whole space into the $R_0<R_{\rm sph}$ and $R_0>R_{\rm sph}$ regions. The pink-filled region and the blue-filled region represent the condition $R_0<R_{\rm sph}$ and $R_0>R_{\rm sph}$, respectively. The dotted line shows the accretion rate $\dot m$ of M82 X-2.}}
		\label{fig:B_mdot}
	\end{figure}
	
	{Combining Equations~(\ref{eq:solve_omega_geq}) and (\ref{eq:B_geq}), and (\ref{eq:solve_omega_leq}) and (\ref{eq:B_leq}), we can get the $B-\dot{m}$ relation. Taking $M_{1.4}=1$, $R_6=1$, $\dot{P}_{-10}=-2$ and $P=1.37\,{\rm s}$ for M82 X-2, we plot the the relation in Figure~\ref{fig:B_mdot} for cases~1 (left panel) and~2 (right panel). In Figure~\ref{fig:B_mdot} the solid curve and two horizontal lines correspond to the $B-\dot m$ relation for $R_0>R_{\rm sph}$ and $R_0<R_{\rm sph}$ respectively. We use different colors to represent different values of the fastness parameter $\omega$. For a given accretion rate $\dot{m}$, there are up to four solutions of $B$ corresponding to the four solutions of $\omega$ from Equations~(\ref{eq:solve_omega_geq}) and (\ref{eq:solve_omega_leq}), but only up to two of them are physical, because the solutions on the curve are only valid in the region filled in blue ($R_0>R_{\rm sph}$) and those on the horizontal lines are only valid in the region filled in pink ($R_0<R_{\rm sph}$). The dividing line between these two regions is drawn with the black dashed line. Please refer to Appendix~\ref{section:appendix} for more detailed derivation.}

	\begin{table}[ht!]
		\caption{\small The spin period $P$, spin period derivation $\dot{P}$, luminosity $L$, mass transfer rate $\dot m$ (in units of $\dot{M}_{\rm Edd}$), the beaming factor $b$, the co-rotation radius $R_{\rm c}$ and the spherical radius $R_{\rm sph}$ for the eight ULX pulsars.\label{table:parameters}}
		\setlength{\tabcolsep}{5pt}
		\small
		\begin{tabular}{llllllll}
			\hline\noalign{\smallskip}
			Sources		&$P\,(\rm s)$		&$\dot{P}\,({\rm s\ s^{-1}})$&$L\,(\rm{erg\ s^{-1}})$&$\dot{m}$&$b$&$R_{\rm c}\,({\rm cm})$&$R_{\rm sph}\,({\rm cm})$\\
			\hline\noalign{\smallskip}
			M82 X-2	$^{[1]}$		&$1.37$	&$-2\times{10}^{-10}$	&$1.8\times{10}^{40}$	&$36.4$&$0.06$&$2.07\times10^{8}$&$7.53\times 10^{7}$\\
			NGC5907 ULX-1$^{[2]}$	&$1.14$&$-8.1\times{10}^{-10}$	&$1.1\times{10}^{41}$	&$82.9$&$0.01$&$1.83\times10^{8}$&$1.71\times 10^{8}$\\
			M51 ULX-7	$^{[3]}$	&$3.0$	&$-2.6\times{10}^{-9}$	&$7.1\times{10}^{39}$	&$24.0$&$0.13$&$3.49\times10^{8}$&$4.96\times 10^{7}$\\
			NGC7793 P13	$^{[4]}$	&$0.42$	&$-4\times{10}^{-11}$	&$5\times{10}^{39}$		&$20.5$&$0.17$&$9.40\times10^{7}$&$4.24\times 10^{7}$\\
			NGC300 ULX-1$^{[5]}$&$\sim 31.6$&$-5.56\times 10^{-7}$	&$4.7\times 10^{39}$	&$20.0$&$0.18$&$1.68\times10^{9}$&$4.13\times 10^{7}$\\
			SMC X-3$^{[6]}$		&$\sim 7.78$&$-7.4\times 10^{-10}$	&$2.5\times 10^{39}$	&$15.1$&$0.32$&$6.58\times10^{8}$&$3.12\times 10^{7}$\\
			NGC2403 ULX$^{[7]}$	&$\sim 18$	&$-1.1\times 10^{-7}$	&$1.2\times 10^{39}$	&$10.9$&$0.61$&$1.15\times10^{9}$&$2.26\times 10^{7}$\\
			Swift J0243.6+6124$^{[8]}$&$9.86$&$-2.1\times 10^{-8}$	&$\gtrsim1.5\times 10^{39}$&$12.1$&$0.50$&$7.71\times10^{8}$&$2.49\times 10^{7}$\\
			\noalign{\smallskip}\hline
		\end{tabular}
		\tablecomments{\textwidth}{
			$^{[1]}$ Donor star mass $> 5\,\rm{M_\odot}$, \cite{Bachetti+m82x2+2014+nature};
			$^{[2]}$ Hyper-luminous sources, \cite{Israel+2017+NGC5907ulx1};
			$^{[3]}$ High mass X-ray binary, \cite{M51ulx7+Rodrguez+2020ApJ,M51ULX7+Vasilopoulos+2020};
			$^{[4]}$ B9Ia supergiant companion, \cite{NGC7793p13+Furst+2016,Furst+2018,Israel+2017+NGC7793P13};
			$^{[5]}$ Be X-ray transient source, \cite{Carpano+2018};
			$^{[6]}$ Be X-ray transient source, \cite{Tsygankov+2017,Townsend+2017};
			$^{[7]}$ Be X-ray transient source, \cite{Trudolyubov+2007ApJ};
			$^{[8]}$ Galactic Be X-ray transient source, \cite{Doroshenko+swift+2018,Swift+2018+distance}.}
	\end{table} 
 
	In Table~\ref{table:results}, we list the calculated strengths of the surface magnetic field of the NS in units of $10^{10}\,{\rm G}$ and $10^{13}\,{\rm G}$ for the low- and the high-$B$ solutions respectively, and we also mark the solutions whether in the situation $R_0>R_{\rm sph}$ or $R_0< R_{\rm sph}$ with notations `$>$' and `$<$' respectively. In addition, the parameter $\xi$ is also listed and is in the range of $0.52-0.91$. Generally a larger $\omega$ ($\gtrsim 0.6$), a stronger magnetic field{, which can be seen from Figure~\ref{fig:B_mdot}}. Most of the low-$B$ solutions fall in the range $R_0\leq R_{\rm sph}$ except NGC2403 ULX, while the high-$B$ solutions fall in the range $R_0\geq R_{\rm sph}$ except NGC5907 ULX-1. We also find that, the fastness parameter $\omega$, $\xi$ and the magnetic field $B$ are the same in cases 1 and 2 for the low-$B$ solutions, because when $\omega\to 0$, $n(\omega)$ in both cases recovers to the same value, i.e., $5/3$. In the following we discuss the sources individually, and compare our results with previous studies.

	\begin{table}[ht!]
		\caption{\small Derived parameters of the eight ULX pulsars. Each solution contains the magnetic field $B$, the fastness parameter $\omega$ and whether the inner radius of disk $R_0$ is greater (`$>$') than $R_{\rm sph}$ or not (`$<$'). The magnetic field of the low-$B$ solutions are in units of $10^{10}\,{\rm G}$ and $10^{13}\,{\rm G}$ for the low- and high-$B$ solutions, respectively. We also list the reference values from other works.\label{table:results}}
		\setlength{\tabcolsep}{1pt}
		\small
		\begin{tabular}{l|c|cccc|cccc|c}
		\hline\noalign{\smallskip}
		\multirow{2}{*}{Sources}&\multirow{2}{*}{Cases}&\multicolumn{4}{c|}{Low-$B$}
		&\multicolumn{4}{c|}{High-$B$}&\multirow{2}{*}{Ref-$B_{13}$}\\
		\noalign{\smallskip}\cline{3-10}\noalign{\smallskip}
		&&{$R_0\,{\rm vs.}\,R_{\rm sph}$}&$\omega$&$\xi$&$B_{10}$	&$R_0\,{\rm vs.}\,\,R_{\rm sph}$&$\omega$&$\xi$&$B_{13}$&\\ 
		\noalign{\smallskip}\hline\noalign{\smallskip}
		\multirow{2}{*}{M82 X-2}	&1&$<$&$0.014$&$0.90$&$4.08$	&$>$&$0.705$&$0.64$&$1.74$
		&\multirow{2}{*}{\begin{tabular}[c]{@{}l@{}}$\lesssim 1^{[1]}$, $\ \gtrsim 1^{[2]}$\\$2-6.7^{[3]}$\end{tabular}}\\
									&2&$<$&$0.014$&$0.90$&$4.07$	&$>$&$0.846$&$0.53$&$2.98$&\\
		\noalign{\smallskip}\hline\noalign{\smallskip}
		\multirow{2}{*}{NGC5907 ULX-1}	&1&$<$&$0.105$&$0.88$&$64.5$	&$<$&$0.681$&$0.65$&$1.78$
		&\multirow{2}{*}{$0.2-3^{[4]}$}\\
										&2&$<$&$0.105$&$0.88$&$64.1$	&$<$&$0.831$&$0.54$&$3.30$&\\
		\noalign{\smallskip}\hline\noalign{\smallskip}
		\multirow{2}{*}{M51 ULX-7}	&1&$<$&$0.018$&$0.90$&$18.3$	&$>$&$0.682$&$0.65$&$3.27$
		&\multirow{2}{*}{\begin{tabular}[c]{@{}c@{}}$0.1-10^{[5]}$\\ $2-7^{[6]}$\end{tabular}}\\
									&2&$<$&$0.018$&$0.90$&$18.3$	&$>$&$0.829$&$0.55$&$5.59$&\\
		\noalign{\smallskip}\hline\noalign{\smallskip}
		\multirow{2}{*}{NGC7793 P13}	&1&$<$&$0.104$&$0.88$&$14.1$	&$>$&$0.651$&$0.67$&$0.275$
		&\multirow{2}{*}{$\sim 0.2^{[7]},\ 0.15^{[8]}$}\\
										&2&$<$&$0.103$&$0.88$&$14.0$	&$>$&$0.805$&$0.57$&$0.471$&\\
		\noalign{\smallskip}\hline\noalign{\smallskip}
		\multirow{2}{*}{NGC300 ULX-1}	&1&$<$&$0.003$&$0.90$&$48.1$	&$>$&$0.680$&$0.65$&$46.2$
		&\multirow{2}{*}{$\sim 0.1^{[9]},\ \gtrsim 1^{[10]}$}\\
										&2&$<$&$0.003$&$0.90$&$48.1$	&$>$&$0.827$&$0.55$&$79.0$&\\
		\noalign{\smallskip}\hline\noalign{\smallskip}
		\multirow{2}{*}{SMC X-3}	&1&$<$&$0.0003$&$0.91$&$0.156$	&$>$&$0.713$&$0.63$&$8.72$
		&\multirow{2}{*}{\begin{tabular}[c]{@{}c@{}}$2-3$\\ or $0.1-0.5^{[11]}$\end{tabular}}\\ 
									&2&$<$&$0.0003$&$0.91$&$0.156$	&$>$&$0.852$&$0.52$&$14.9$&\\
		\noalign{\smallskip}\hline\noalign{\smallskip}
		\multirow{2}{*}{NGC2403 ULX	}	&1&$>$&$0.005$&$0.90$&$36.4$	&$>$&$0.666$&$0.66$&$16.9$
		&\multirow{2}{*}{$-$}\\
										&2&$>$&$0.005$&$0.90$&$36.3$	&$>$&$0.817$&$0.56$&$29.0$&\\
		\noalign{\smallskip}\hline\noalign{\smallskip}
		\multirow{2}{*}{Swift J0243.6+6124}	&1&$<$&$0.004$&$0.90$&$11.6$	&$>$&$0.682$&$0.65$&$9.28$
		&\multirow{2}{*}{\begin{tabular}[c]{@{}c@{}}{$<1$$^{[12]}$, $\gtrsim 1^{[12]}$}\\$\sim 0.1^{[13]},\ \gtrsim 2.4^{[14]}$\end{tabular}}\\ 
											&2&$<$&$0.004$&$0.90$&$11.6$	&$>$&$0.829$&$0.55$&$15.9$&\\
		\noalign{\smallskip}\hline\noalign{\smallskip}
		\end{tabular}
		\tablecomments{\textwidth}{
			$^{[1]}$\cite{Xu+Li+2017+M82x-2}; 
			$^{[2]}$\cite{Bachetti+m82x2+2014+nature}; 
			$^{[3]}$\cite{M82X-2+Eksi+2015}; 
			$^{[4]}$\cite{Israel+2017+NGC5907ulx1}; 
			$^{[5]}$\cite{M51ulx7+Rodrguez+2020ApJ}; 
			$^{[6]}$\cite{M51ULX7+Vasilopoulos+2020}; 
			$^{[7]}$\cite{Israel+2017+NGC7793P13}; 
			$^{[8]}$\cite{NGC7793p13+Furst+2016}; 
			$^{[9]}$ Measured by the CRSF, \cite{Walton+2018+ApJ+cyc+ngc300}; 
			$^{[10]}$\cite{Koliopanos+2017+A&A+AEmodel}; 
			$^{[11]}$\cite{Tsygankov+2017}; 
			$^{[12]}$\cite{Tsygankov+2018MNRAS.479L.134T};
			$^{[13]}$\cite{Doroshenko+swift+2018};
			$^{[14]}$\cite{Kong_2020ApJ...902...18K}.}
	\end{table}

	\subsection{M82 X-2}
	
	\hspace{15pt}M82 X-2 is the first discovered ULXs powered by an accreting NS (\citealt{Bachetti+m82x2+2014+nature}). \cite{Xu+Li+2017+M82x-2} considered the thin and thick disk models and reported that $B\lesssim 10^{13}\,{\rm G}$. \cite{Bachetti+m82x2+2014+nature} pointed out that, to maintain an accreting gas column, there should be a strong enough magnetic field $B\geq10^{13}\,{\rm G}$ (\citealt{Basko+1976MNRAS.175..395B}), and even a stronger field $B\sim 10^{14}\,{\rm G}$ could be plausible due to the reduction of the electron scattering opacity. \cite{M82X-2+Eksi+2015} showed that the dipole magnetic field is at least $2\times 10^{13}\,\rm G$ and even $6.7\times 10^{13}\,\rm G$, exceeding the quantum critical magnetic field $B_{\rm c}=4.4\times 10^{13}\,\rm G$, derived from a simplified dimensionless torque $n=1-\omega/\omega_{\rm crit}$ with $R_0=0.5R_{\rm A}$.  In our models, the high-$B$ solutions give $B= 1.74\times10^{13}\,{\rm G}$ and $2.98\times10^{13}\,{\rm G}$ in cases~1 and~2, respectively, while the low-$B$ solutions give a weaker magnetic field of $B\sim4\times 10^{10}\,{\rm G}$ for both cases~1 and~2.  Based on the observed luminosity, the high-$B$ solutions could be more realistic, because a strong magnetic field can reduce the electron scattering cross section $\sigma_{\rm T}$ (\citealt{Canuto+thomson+scattering+1971+PhRvD}), and enhance the Eddington luminosity $L_{\rm Edd}$.
	
	\cite{DallOsso+etl+2015} adopted a torque model from \cite{GL79a,GL79b} to study the magnetic field of M82 X-2. They pointed out that the low-$B$ branch is far from the state of spin equilibrium, corresponding to the Alfv\'en radius $R_{\rm A}<{10}^7\,{\rm cm}$. In case~1, we have $R_0=\omega^{2/3}R_c=12.2\, R_{\rm NS}$, and $R_{\rm c}=207\, R_{\rm NS}$, corresponding to the fastness parameter $\omega=0.014$, and the dimensionless torque $n(\omega)=1.65$ which is 20 times greater than $n|_{\omega=0.705}=0.07$ of the high-$B$ solutions under the same conditions. The inner radius of the disk $R_0=\xi R_{\rm A}$ is very close to the surface of NS, and the spin-down torque generated by the accretion disk outside the co-rotation radius $R_{\rm c}$ can be ignored. However, the high-$B$ solution is close to the spin equilibrium where the inner radius of the disk is far away from the surface of the NS. Therefore, the accretion disk outside $R_{\rm c}$ can bring a large reverse torque and reduce the total torque on the NS.
	
	Recently, \cite{Bachetti+2020ApJ...891...44B} analyzed the timing behaviour of M82 X-2 and obtained an average spin-down rate $\dot{\nu}=-\dot{P}/P^2\sim -6\times 10^{-11}\,{\rm Hz\ s^{-1}}$ between 2014 and 2016, in contrast with the strong spin-up rate during the 2014 observations (\citealt{Bachetti+m82x2+2014+nature}). \cite{Bachetti+2020ApJ...891...44B} pointed out that M82 X-2 is close to the spin equilibrium, because M82 X-2 alternates between the spin-up and spin-down. In our models, assuming that M82 X-2 is in spin equilibrium state, it leads to the magnetic fields $B=1.79\times10^{13}\,{\rm G}$ and $3.08\times 10^{13}\,{\rm G}$ corresponding to $\omega_{\rm crit}=0.714$ and $0.853$ for cases~1 and~2, respectively.
	
	\subsection{NGC5907 ULX-1}
	
	\hspace{15pt}In our model, the low-$B$ solutions give $B\sim 6\times10^{10}\,{\rm G}$ in both cases~1 and~2, and the high-$B$ solutions give $B=1.78\times10^{13}\,{\rm G}$ and $3.30\times10^{13}\,{\rm G}$ in cases~1 and~2 respectively with a beaming factor $b=0.01$. \cite{Israel+2017+NGC5907ulx1} found that NGC5907 ULX-1 is a hyper-luminous ULX pulsar (with $L\geq 10^{41}\,{\rm erg\ s^{-1}}$) and pointed out that, a multipolar magnetic field at the NS surface of $B_{\rm multi}\sim (0.7-3)\times 10^{14}\,{\rm G}$ together with a $(0.2-3)\times10^{13}\,{\rm G}$ dipole component and a beaming factor $b\sim 1/25-1/7$ are necessary to interpret the properties of NGC5907 ULX-1. 
	
	\subsection{M51 ULX-7}
	
	\hspace{15pt}In our model, the high-$B$ solutions give $B=3.27\times 10^{13}\,{\rm G}$ and $5.59\times10^{13}\,{\rm G}$ in cases~1 and~2 respectively with a beaming factor $b=0.13$. \cite{M51ULX7+Vasilopoulos+2020} used the standard accretion model (\citealt{GL79a,GL79b,Wang+1995}) to explain the properties of M51 ULX-7 and obtained a surface magnetic field of $(2-7)\times 10^{13}\,{\rm G}$ assuming that the NS was near spin equilibrium. They also analyzed the X-ray light curve and suggested that, if $39\,{\rm d}$ super-orbital period results from the precession of the NS, it may imply a surface magnetic field of $(3-4)\times 10^{13}\,{\rm G}$ assuming that the distortion ($\varepsilon$) of the NS relies on the surface magnetic field energy ($\varepsilon \propto B^2$). \cite{M51ulx7+Rodrguez+2020ApJ} suggested that M51 ULX-7 may have a massive OB giant or supergiant donor and its dipole magnetic field is $(0.1-10)\times 10^{13}\,{\rm G}$ with a weakly beamed emission $b\sim 1/12-1/4$. They also pointed out that a stronger multipolar component ($\sim 10^{14}\,{\rm G}$) at the surface of the NS could not be excluded.

	\subsection{NGC7793 P13}
	
	\hspace{15pt}\cite{NGC7793p13+Furst+2016} reported the detection of $\sim 0.42 \,{\rm s}$ pulsations from NGC7793 P13 and estimated its magnetic field of $B\simeq 1.5\times 10^{12}\,{\rm G}$ using the standard accretion disk models (\citealt{GL79a}). They also pointed out that a high-degree of beaming could account for the ultra-high luminosity. \cite{Israel+2017+NGC7793P13} obtained a surface dipole field of $B\sim 2\times 10^{12}\,{\rm G}$ assuming a maximum accretion luminosity of $\sim 10^{39}\,{\rm erg\ s^{-1}}$ with a beaming factor of $b\sim 1/15$. A multipolar magnetic field of $B_{\rm multi}>8\times 10^{13}\,{\rm G}$ at the base of the accretion column was also obtained to make the maximum accretion luminosity of $9\times 10^{39}\,{\rm erg\ s^{-1}}$ possible. \cite{Israel+2017+NGC7793P13}  pointed out that the magnetic field is dominated by multipole component at the surface of NS, but by the dipole component close to the magnetospheric radius $R_{\rm A}$, due to the weaker steepness of dipole magnetic field with respect to radius ($\sim R^{-3}$, for example, compared with $\sim R^{-5}$ for quadrupole field). Our high-$B$ solutions are in accordance with the dipole components magnetic field estimated by others: the high-$B$ solutions give $B=2.75\times10^{12}\,{\rm G}$ and $4.71\times10^{12}\,{\rm G}$ in cases~1 and~2 respectively with a beaming factor $b=0.17$.
	
	\subsection{NGC300 ULX-1}
	\label{subsection}
	\hspace{15pt}Using phase-resolved broadband spectroscopy observed with \textit{XMM-Newton} and \textit{NuSTAR}, \cite{Walton+2018+ApJ+cyc+ngc300} discovered a likely CRSF at $E_{\rm cyc}\sim 13\,\rm{k eV}$ of NGC300 ULX-1, which implies a magnetic field $B\sim 10^{12}\,\rm {G}$ for electron scattering. However, \cite{Koliopanos+2019+A&A+cyc} found that, although the CRSF can be interpreted by a broad Gaussian absorption line with a magnetic field $\sim 10^{12}\,\rm G$, the multicolour accretion envelope model (\citealt{Mushtukov+2017+MNRAS+AEmodel,Koliopanos+2017+A&A+AEmodel}) and a hard power-law tail can also account for the spectral and temporal emission characteristics of NGC300 ULX-1, questioning whether the CRSF exists or not. Our high-$B$ branch solutions give $B=4.62\times 10^{14}\,\rm{G}$ in case~1 and $B=7.9\times 10^{14}\,\rm{G}$ in case 2 which are much stronger than the values predicted by the CRSF and exceed the quantum magnetic limit $B_{\rm crit}$; instead, the low-$B$ solution of $B\sim 4.8\times 10^{11}\,{\rm G}$ in both cases~1 and~2 seem to be close to the values from the CRSF.
	
	Although the spin-up rates $\dot{P}$ of Be-type ULX pulsars are much larger than those of the persistent ones, all of the eight ULX pulsars have similar $\dot\nu=-\dot P/P^2\sim 10^{-10}\,{\rm Hz\ s^{-1}}$ which indicates a similar accretion torque. We notice that NGC300 ULX-1 has a long spin period $\sim 31.6\,{\rm s}$ compared with, for example, M82 X-2 ($P\sim 1.37\,{\rm s}$). Since $\omega\propto P^{-1}$, a longer spin period $P$ corresponds to a smaller fastness parameter $\omega$, so NGC300 ULX-1 may be far from spin equilibrium.
	
	\subsection{SMC X-3}
	
	\hspace{15pt} In our model, the magnetic field of high-$B$ solutions give $B=8.72\times10^{13}\,{\rm G}$ and $14.9\times10^{13}\,{\rm G}$ in cases~1 and~2 respectively while the low-$B$ solutions give $B\sim 1.56\times 10^{9}\,{\rm G}$ corresponding to $\omega\sim 3\times10^{-4}$. From the data observed with \textit{Swift}/XRT, \textit{Fermi}/GBM and \textit{NuSTAR} for SMC X-3, \cite{Tsygankov+2017} found a change in its pulse profile when the luminosity was $(2-3)\times 10^{38}\,{\rm erg \ s^{-1}}$ and argued that the change was caused by the disappearance of the accretion column and hence a variation of the intrinsic X-ray beaming from the pulsar, which indicates a magnetic field of $B\sim(2-3)\times 10^{13}\,{\rm G}$ using a model in \cite{Mushtukov+2015MNRAS.454.2539M}. \cite{Tsygankov+2017} also took into account the propeller effect when the transition luminosity was in the range of $(0.3-70)\times 10^{35}\,{\rm erg\ s^{-1}}$ and estimated the dipole magnetic field of $B\sim (1-5)\times 10^{12}\,{\rm G}$. They pointed out that the two different values of the magnetic field estimated above made SMC X-3 a candidate for ULX pulsars with a significant multipole magnetic field component, and the sources like SMC X-3 may contribute to the intermediate ULX population between classical X-ray pulsars and accreting magnetars.
	
	\subsection{NGC2403 ULX}
	
	\hspace{15pt}NGC2403 ULX (\citealt{Trudolyubov+2007ApJ}) is a transient X-ray pulsar with a peak luminosity exceeding $10^{39}\,{\rm erg \ s^{-1}}$. In our model, the low-$B$ solutions give $B\simeq3.6\times10^{11}\,{\rm G}$ for cases~1 and~2, and the high-$B$ solutions give $B=1.69\times 10^{14}\,{\rm G}$ and $2.9\times10^{14}\,{\rm G}$ for cases~1 and~2 respectively. Similar to NGC300 ULX-1, NGC2403 ULX also has a relatively long spin period $P\sim18\,{\rm s}$. {Considering the fact that the main sequence lifetime ($\sim10^{7}\,{\rm yr}$) of a Be star is much longer than the field decay time ($\sim 10^{3}-10^4\,{\rm yr}$) for a magnetar,} the low-$B$ solutions may be more physical. \cite{King2019+no+magnetar} considered only the material torque and estimated the magnetic field of $B\sim 5.6\times 10^{11}\,{\rm G}$ which is similar to the low-$B$ solutions with the dimensionless total torque $n\simeq 1.66$.
	
	\subsection{Swift J0243.6+6124}
	
	\hspace{15pt}Swift J0243.6+6124 (\citealt{Swift+2018+distance}) is the first discovered ULX candidate in the Milky Way. In our model, the low-$B$ solutions give $B\simeq 1.16\times 10^{11}\,{\rm G}$ for both cases~1 and~2 and the high-$B$ solutions give $B=9.28\times 10^{13}\,{\rm G}$ and $1.59\times 10^{14}\,{\rm G}$ for cases~1 and~2, respectively. \cite{Doroshenko+swift+2018} modelled the spin variations of the NS and obtained a magnetic field of $\sim 10^{12}\,{\rm G}$. \cite{Tsygankov+2018MNRAS.479L.134T} gave an upper limit on the propeller luminosity $\le 6.8\times 10^{35}\,{\rm erg\ s^{-1}}$ which implied a dipole magnetic field component $B<10^{13}\,{\rm G}$, but they also estimated the magnetic field $B\gtrsim 10^{13}\,{\rm G}$ if assuming the variation in the pulse profile was related to a critical luminosity $\sim 3\times 10^{38}\,{\rm erg\ s^{-1}}$ {associated with the onset of the accretion column}. {They concluded that these two independent estimations were marginally compatible if taking the effective magnetosphere size (the parameter $\xi$ in our work) into consideration, and they also emphasized that the transition to the propeller was actually not observed which led to the inconsistency in their two different results, because the transition luminosity could be lower.} Recently, \cite{Kong_2020ApJ...902...18K} analyzed the $1-100\, {\rm keV}$ data observed with the \textit{Hard X-ray Modulation Telescope} (Insight-\textit{HXMT}) during the 2017-2018 outburst. They found a spectral transition at two typical luminosities ($L_1\sim 1.5\times 10^{38}\,{\rm erg\ s^{-1}}$ and $L_2\sim4.4\times 10^{38}\,{\rm erg\ s^{-1}}$) and estimated a magnetic field of $B\sim 2.4\times 10^{13}\,{\rm G}$ related to $L_{2}$.\\
	
	\begin{figure}[h]
		\begin{subfigure}{\textwidth}
			\includegraphics[width=\textwidth]{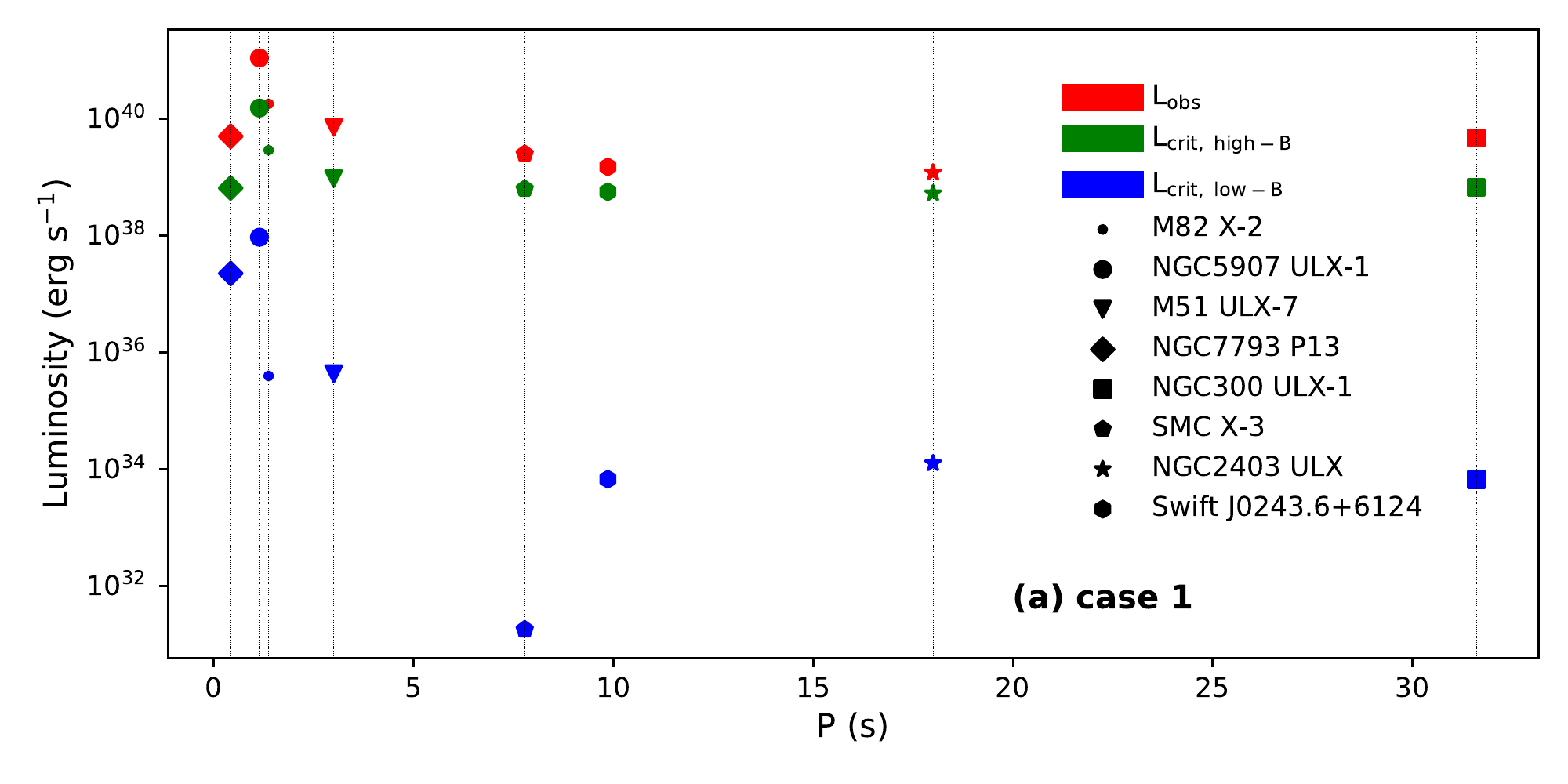}
		\end{subfigure}
		\begin{subfigure}{\textwidth}
			\includegraphics[width=\textwidth]{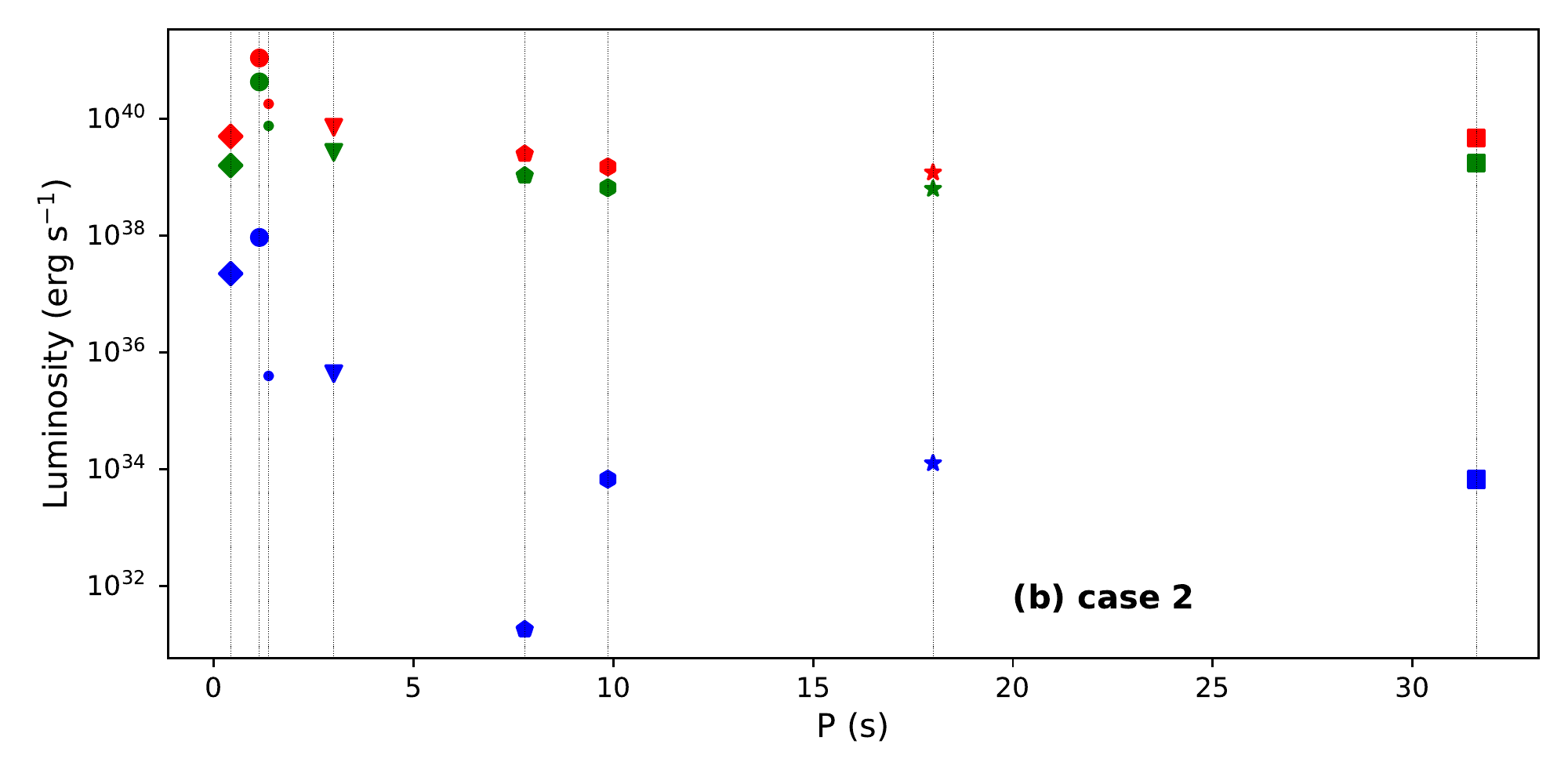}
		\end{subfigure}
		\caption{\small When $R_0=R_{\rm c}$, a transition between accretor and propeller regimes occurs. The critical luminosity $L_{\rm crit}$ are illustrated by blue and green markers corresponding to low-$B$ and high-$B$ magnetic field solutions respectively, and the observed isotropic luminosity $L_{\rm obs}$ are also illustrated by red markers. Panel (a) and (b) represent cases~1 and~2 respectively.}
		\label{fig:transient}
	\end{figure}

	\section{Discussion}
	\label{section:disscussion}

	\subsection{Comparison With \cite{Erkut+2020ApJ...899...97E}}
	
	\hspace{15pt}Both our work and \cite{Erkut+2020ApJ...899...97E} estimate the NS magnetic fields based on the magnetized accretion disk model. So it is interesting to compare the differences between them.
		\begin{enumerate}
			\item[a.] We note that both works use the angular momentum conservation (Equation~\ref{eq:boundary}) to derive the inner disk radius. \cite{Erkut+2020ApJ...899...97E} integrated both sides of Equation~(\ref{eq:boundary}) over the boundary layer. This introduces the width $\delta=\Delta R/R_{0}$ of the boundary layer which is unknown and assumed to between $0.01$ and $0.3$. In this case, the inner radius of the disk is $R_0=2^{1/7}\delta^{2/7}R_{\rm A}\sim (0.30-0.78)R_{\rm A}$, while in our results, $\xi\sim 0.52-0.91$. The problem with this approach is that $B_\phi=\gamma_\phi B_z$ is implicitly assumed to be nearly constant over the boundary layer, although $\Omega$ actually varies drastically from $\Omega_{\rm K}$ to $\Omega_{\rm s}$.
			
			\item[b.] \cite{Erkut+2020ApJ...899...97E} estimated the magnetic fields by solving the torque equation (Equation~\ref{eq:roation}) and assumed the total dimensionless torque $n$ as a constant of order unity, which means that the magnetic torque contributed by the disk outside $R_0$ is ignored and $n$ is always greater than zero. While in our work, the magnetic torque is always non-negligible. Taking M82 X-2 for example, the magnetic torques $N_{\rm mag}$ for high-$B$ solutions are about$-0.927 N_0$ and $-0.931N_0$ for cases~1 and~2, respectively, whose absolute values are comparable with $N_0$. If we set $n=1$ in Equation~(\ref{eq:P,Pdot,omega}), we have $B\simeq8.73\times 10^{10}\,{\rm G}$ ($\ll 10^{13}\,{\rm G}$) for M82 X-2. These results are similar to the low-$B$ solutions and \cite{King2019+no+magnetar} made a similar estimation of the ULX pulsars' magnetic fields ($B\lesssim 10^{11}\,{\rm G}$) without considering the magnetic torque. However, the spin-up timescale $P/|\dot P|\sim 10^2\,\rm yr$ of the ULX pulsars and the alternation between the spin-up and spin-down observed in M82 X-2 (\citealt{Bachetti+2020ApJ...891...44B}) indicate that the NS must be close to the spin equilibrium with its disk, which conflicts with $n\simeq 1$.

			\item[c.]  \cite{Erkut+2020ApJ...899...97E} assumed that the beaming factor $b$ is roughly the fractional polar cap area, determined by not only the accretion rate but also the magnetic field and the magnetic inclination angle. This leads to the conclusion that the beaming factor increases with the accretion rate, i.e. $b\propto\dot m^{2/7}$, which is inconsistent with traditional picture of the beaming effect. Moreover, for accreting magnetized NSs, the site of radiation is likely the accretion column rather the polar cap. We adopt the empirical relation suggested by \cite{King+2009+ULXs} with $b\propto \dot m^{-2}$.	
		\end{enumerate}
 	While points (a) and (b) probably do not cause significant differences in the final results, points (c) can results in substantially different estimates of the real accretion rates. That is why \cite{Erkut+2020ApJ...899...97E} had to adjust the masses and the radii of specific ULX pulsars.
	
	\subsection{ The Low- and High-$\mathbf B$ Solutions}
	
	\hspace{15pt}As seen from Table~\ref{table:results}, there are low- and high-$B$ solutions for each ULX pulsar from their spin evolution. It is essential to discriminate which one is real. There are several kinds of ways. First, the CRSFs present direct measure of the surface dipole magnetic field, if the multipole component does not dominate. Second, most known ULX pulsars are likely in high-mass X-ray binaries with typical ages $\sim 10^6-10^7\,{\rm yr}$, so a magnetar's field seems unlikely. A comparison with Galactic X-ray pulsars in HMXBs also indicate a magnetic field of $10^{11}-10^{13}\,{\rm G}$. In addition, there is another way to discriminate the high- and low-$B$ solutions. If the X-ray luminosities experience a large change, the NS may transit between accretor and propeller regimes. {Assuming that $\xi$ is a constant}, the condition of this change is $R_0=R_{\rm c}$  and the critical accretion rate for this transition is
	\begin{equation}
		\dot{m}_{\rm in,crit}=\omega^{7/3}\dot{m}_{\rm in},
	\end{equation} 
	where $\dot{m}_{\rm in,crit}$ denotes the critical mass accretion rate at the inner radius of the disk in units of $\dot{M}_{\rm Edd}$ and $\dot{m}_{\rm in}$ is the current accretion rate at the inner radius of the disk, $\omega$ is the current fastness parameter. Actually, in this transitional critical state, the inner radius $R_0$ of the disk is always greater than the spherization radius $R_{\rm sph}$, so, $\dot m_{\rm crit}=\dot m_{\rm in,crit}$, because the co-rotation radius $R_{\rm c}$ is always greater than the spherization radius $R_{\rm sph}$ for the eight ULX pulsars. We calculate their critical luminosity $L_{\rm crit}$ using Equation~(\ref{eq:L_mdot}) in both cases~1 and~2 for the two branches of solutions. In Figure~\ref{fig:transient}, we illustrate $L_{\rm crit}$ and $L_{\rm obs}$ for the eight ULX pulsars. Obviously the ULX pulsars with high-$B$ fields are more likely to experience transitions. Taking the high-$B$ magnetic field of the transient source SMC X-3 for example, we find that $\dot{m}_{\rm crit}=6.85$ and $L_{\rm crit}=6.31\times 10^{38}\,{\rm erg\ s^{-1}}$, but $\dot{m}_{\rm crit}=8.29\times 10^{-8}$ and $L_{\rm crit}=1.79\times 10^{31}\,{\rm erg\ s^{-1}}$ for the low-$B$ magnetic field solutions in case~1.
	
	\subsection{The Be-type ULX Pulsars}
	
	\hspace{15pt}While the high-$B$ solutions for the four persistent ULX pulsars are consistent with other works, we find that the derived magnetic fields of Be-type ULX pulsars are either too high or too low. \cite{Klus+2014MNRAS.437.3863K} analyzed the long-term average spin change rates and the average X-ray luminosity of 42 Be/X-ray binaries in the Small Magellanic Cloud, and reached a conclusion that a large fraction of the NSs likely have magnetic fields $\gtrsim 10^{14}\,{\rm G}$ or $\sim 10^{6}-10^{10}\,{\rm G}$ assuming that the NSs are close to or far away from the spin equilibrium, respectively. Both of the derived magnetic fields disagree with the magnetic fields ($10^{11}-10^{13}\,{\rm G}$) measured by the CRSFs in Galactic Be X-ray binaries (see, fig.~8 in \citealt{Klus+2014MNRAS.437.3863K}). We encounter the same problem in explaining the low- and high-$B$ solutions of Be-type ULX pulsars. We note that the NSs in Be/X-ray binaries usually capture material from the circumstellar disk of the companion star only at periastron, which leads to X-ray outbursts and transient characteristics, while during the quiescent phase, the accretion disk  may become  advection-dominated (\citealt{Okazaki+2013PASJ...65...41O}) with very low luminosities ($\sim10^{33}\,{\rm erg\ s^{-1}}$, \citealt{Yang+2017ApJ...839..119Y}). This means that the spin evolution of the NSs in most Be/X-ray binaries is determined by a combination of the spin-up torque during outbursts and the spin-down torque during quiescence (\citealt{Xuxiaotian+2019ApJ...872..102X}). As a result, the average spin evolutions and the peak luminosities during outbursts may not provide adequate estimates of the NSs' magnetic fields in Be-type ULX pulsars.
	
	\subsection{The Beaming Factor $b$}
	
	\hspace{15pt}Another issue to be addressed is whether the X-ray radiation is isotropic for ULX pulsars. We have adopted the beaming prescription suggested by \cite{King+2009+ULXs} for super-Eddington accretion. However, the sinusoidal pulse profiles discovered in, for example, M82 X-2, NGC5907 ULX-1, NGC7793 P13 indicate that ULX pulsars probably do not have a strong beaming. \cite{Mushtukov+2021MNRAS.501.2424M} performed Monte Carlo simulations to trace the photons emission and pointed out that the geometrical beaming models may not be consistent with the observations of a large pulsed fraction and the apparent luminosity may be close to their true luminosity. If that is the case, involving beaming factors to infer the accretion rate may be problematic, {and a completely new model is needed for super-Eddington accretion disks}.\\
	
	Finally we summarize our work as follows. We use the super-Eddington, magnetized accretion disk model to calculate the magnetic field strengths of eight ULX pulsars from their observed spin-up variations and luminosities. We obtain two branches of solutions of the magnetic fields distributed in the range of $B\sim (0.156-64.5)\times 10^{10}\,{\rm G}$ and $B\sim (0.275-79.0)\times 10^{13}\,{\rm G}$. The low-$B$ solutions correspond to the state that the NS is far away form the spin equilibrium state, and the high magnetic field case is close to the spin equilibrium. Since direct detection of the CRSFs is still lacking, we suggest a possible way to discriminate the high- and low-$B$ solutions by means of the transition between the accretion and the propeller regimes. We also notice that the magnetic fields of the persistent ULX pulsars are consistent with other works, while the magnetic fields of the Be-type ULX pulsars are not, under the assumption that they are accreting via accretion disk. This implies that the accretion model in Be-type ULX pulsars is likely more complicated than in persistent ULX pulsars.

	\normalem
	\begin{acknowledgements}
	We are grateful to an anonymous referee for his/her valuable comments. This work was supported by the National Key Research and Development Program of China (2016YFA0400803), the Natural Science Foundation of China under grant No. 11773015, 10241301, and Project U1838201 supported by NSFC and CAS.
	\end{acknowledgements}

	\bibliographystyle{raa}
	\bibliography{ms2020-0471.bib}
	
	\appendix
	\section{The Derivation of $B$\lowercase{$-\dot m$} relation}
	\label{section:appendix}
	If $R_0>R_{\rm sph}$, using Equations~(\ref{eq:Mdotin}) and (\ref{eq:roation}), we have
	\begin{equation}
		\dot m=-2.48M_{1.4}^{1/3}R_6^2\dot{P}_{-10}P_1^{-7/3}[n(\omega)]^{-1}\omega^{-1/3},
		\label{eq:mu-mdot}
	\end{equation}
	and, using Equations~(\ref{eq:R_0}) and (\ref{eq:mu-mdot}) and taking $\xi=2^{-1/7}(1-\omega)^{2/7}$, we can obtain
	\begin{equation}
		B=2.57\times10^{12} M_{1.4}R_6^{-2}|\dot P_{-10}|^{1/2}(1-\omega)^{-1/2}\omega [n(\omega)]^{-1/2}\,{\rm G}.
	\end{equation}
	 Combining these two equations for $\omega\in((R_{\rm NS}/R_{\rm c})^{3/2},\ \omega_{\rm crit})$ gives the solid curve for the $B-\dot m$ relation in Figure~\ref{fig:B_mdot}.	
	If $R_0<R_{\rm sph}$, Equation~(\ref{eq:solve_omega_leq}) no longer contains $\dot{m}$, we can get the values of $\omega$ by solving Equation~(\ref{eq:solve_omega_leq}), and then submit into Equation~(\ref{eq:B_leq}) to get the values of the magnetic fields $B$. This leads to the two horizontal lines in Figure~\ref{fig:B_mdot}.	
	The dividing line (the black dashed line in each panel of Figure~\ref{fig:B_mdot}) between the $R_0>R_{\rm sph}$ and $R_0<R_{\rm sph}$ regions is derived as follows, using Equation~(\ref{eq:R_0}), and
	 $$R_{\rm sph}=\frac{3GM\dot{M}}{2L_{\rm Edd}}$$
	 $$R_0=\omega^{2/3} R_{\rm c}=R_{\rm sph},$$ 
	 we can obtain that
	\begin{equation}
		\dot{m}=81.06M_{1.4}^{-2/3}P_1^{2/3}\omega^{2/3},
	\end{equation}
	and
	\begin{equation}
		B=1.47\times10^{13}R_6^{-3}M_{1.4}^{1/2}P_1^{3/2}(1-\omega)^{-1/2}\omega^{3/2}\,{\rm G}.
	\end{equation}

\end{document}